\documentclass{iopjournal}
\usepackage{lmodern}
\usepackage{url}
\usepackage{tabularx}
\usepackage{booktabs}
\usepackage{makecell}
\usepackage{multirow}
\usepackage{pifont} % for checkmarks
\newcommand{\cmark}{\ding{51}}
\usepackage{newtxtext}
\usepackage{newtxmath}
\usepackage{subcaption}

\begin{document}

% \articletype{Perspective} %	 e.g. Paper, Letter, Topical Review...

% \newcommand{\myrev}[1]{{\color{blue}#1}}
% \newcommand{\myadd}[1]{{\color{red}#1}}
\newcommand{\myrev}[1]{{\color{black}#1}}
\newcommand{\myadd}[1]{{\color{black}#1}}

\title{Generative and multimodal AI for materials prediction and design: Progress, challenges, and perspectives}

% \author{Author Name$^1$\orcid{0000-0000-0000-0000}, Author Name$^2$\orcid{0000-0000-0000-0000} and Author Name$^{1,*}$\orcid{0000-0000-0000-0000}}

% \author{Anonymous}

\author{Xianyuan Liu$^{1,2}$, 
Charles Anjah$^{1,2}$,
Benjamin E. Jolly$^{3,4}$, 
Jonathon F. S. Markanday$^{5}$, 
Joshua Berry$^{3}$,
Haolin Wang$^{1,2}$,
Nicola A. Morley$^{3}$,
Robert D. J. Oliver$^{3}$,
Alexandra J. Ramadan$^{6}$,
Delvin Ce Zhang$^{1}$,
Katerina A. Christofidou$^{3,4,*}$,
Haiping Lu$^{1,2,*}$
}

\affil{$^1$School of Computer Science, University of Sheffield, Sheffield, UK}

\affil{$^2$Centre for Machine Intelligence, University of Sheffield, Sheffield, UK}

\affil{$^3$School of Chemical, Materials and Biological Engineering, University of Sheffield, Sheffield, UK}

\affil{$^4$Henry Royce Institute, Royce Discovery Centre, University of Sheffield, Sheffield, UK}

\affil{$^5$Materials Nexus Ltd., Salisbury House, Cambridge, UK}

\affil{$^6$School of Mathematical and Physical Sciences, University of Sheffield, Sheffield, UK}

\affil{$^*$Author to whom any correspondence should be addressed.}

\email{k.christofidou@sheffield.ac.uk; h.lu@sheffield.ac.uk}


%-----------%
\begin{abstract}

\myrev{Artificial intelligence (AI) is accelerating materials prediction and design by enabling efficient exploration of chemical and structural spaces, with particular promise for novel materials discovery. However, novelty in materials discovery encompasses chemical plausibility, structural distinctiveness, property relevance and experimental realisability, making AI-driven novelty claims difficult to substantiate. We introduce a materials property hierarchy, from intrinsic, composition-determined properties to extrinsic, processing-dependent performance, to clarify deployment constraints and distinguish structural, physical and deployment novelty. This framework motivates an evidence-based view of multimodal materials data spanning chemical composition, microstructure, processing, and testing and characterisation, showing that current evidence remains concentrated in composition and idealised structure while heterogeneous, under-represented and weakly integrated modalities limit support for physical and deployment novelty. It also highlights the limitations of benchmarks based mainly on computational labels and proxy novelty criteria. Community-wide standards for data collection, modality alignment and evidence synthesis are needed to support multimodal data construction, process-aware multimodal modelling, feasibility-first generative modelling and deployment-aware benchmarking, so that generative and multimodal AI can design experimentally realisable materials with defensible scientific and practical novelty.}

\end{abstract}

%-----------%
\section{Introduction}
\begin{figure}[t]
\centering
\begin{subfigure}{0.47\linewidth}
    \centering
    \includegraphics[width=\linewidth]{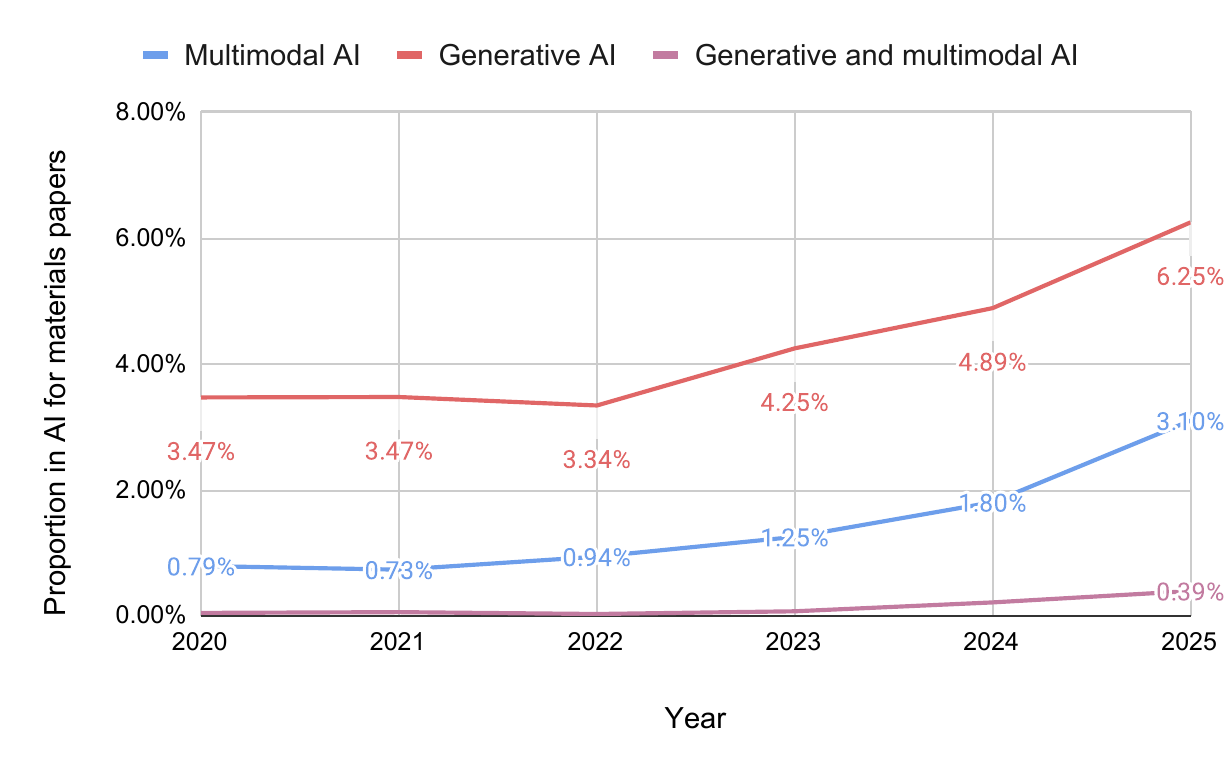}
    \caption{}
    \label{fig:proportion}
\end{subfigure}
\hfill
\begin{subfigure}{0.47\linewidth}
    \centering
    \includegraphics[width=\linewidth]{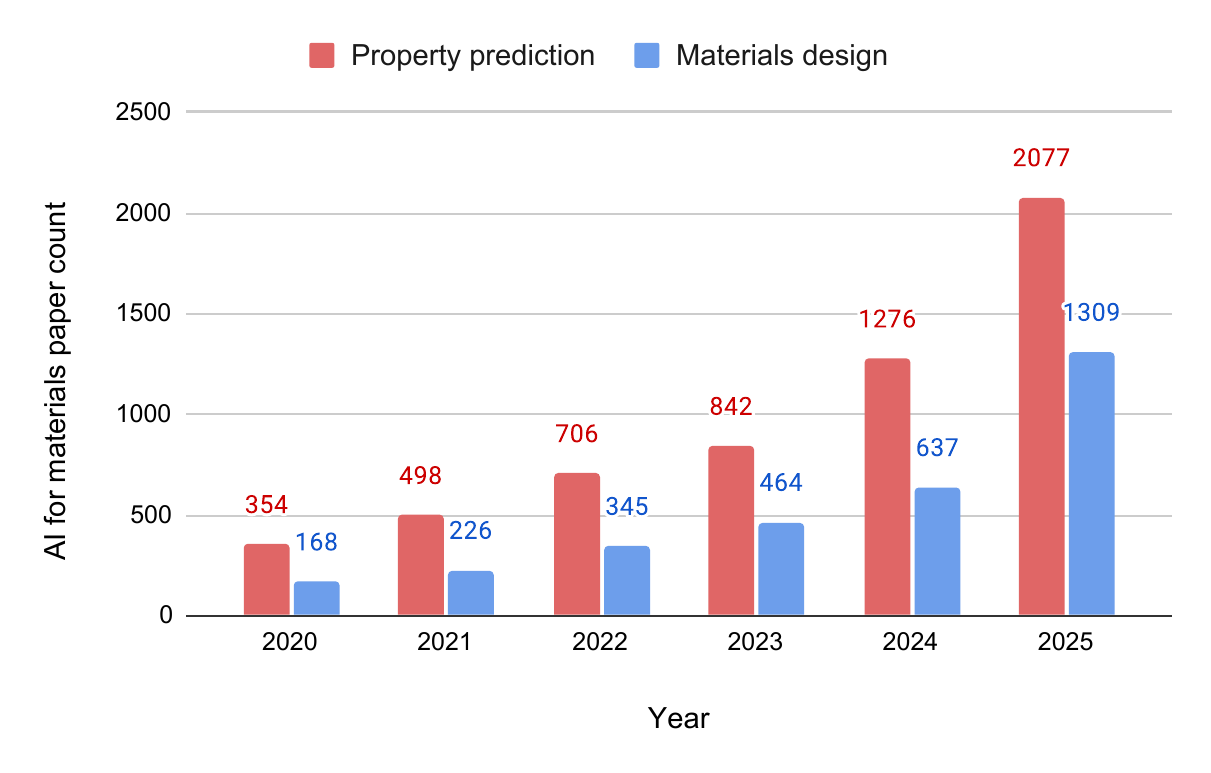}
    \caption{}
    \label{fig:tasks}
\end{subfigure}
\caption{\myrev{Growth trends of AI for materials research from 2020 to 2025. (a) The annual proportion of multimodal AI, generative AI, and generative and multimodal AI publications relative to all AI for materials publications. The ``generative and multimodal AI'' term refers to publications that matched both the ``multimodal AI'' and ``generative AI'' groups. (b) The annual number of AI for materials publications in two key tasks, property prediction and materials design. Publication data were retrieved from the Elsevier Scopus database on 24 June 2026. See Supplementary Information Section 1 for details of the trend analysis.}}
\label{fig:scopus-mmai}
\end{figure}

Artificial intelligence (AI) has emerged as a practical engine for materials research, enabling the design and optimisation of candidate systems by navigating vast, complex compositional spaces to support exploration beyond manual reasoning or trial-and-error~\cite{merchant2023scaling,zeni2025generative}.
Figure \ref{fig:scopus-mmai} shows that generative and multimodal AI have recently expanded rapidly in the materials research domain, both in property prediction and materials design. 

Defining novelty in materials discovery remains challenging. Novelty is established not solely from unfamiliar composition or structure, but also from chemical plausibility, physical behaviour, and experimental realisability. The criteria for establishing novelty depend on the design objective, and the evidence required to support novelty is distributed across materials data. Materials data are inherently multimodal, spanning chemical composition, crystalline phase, process history, characterisation data, and the technical narratives contained in laboratory records and the scientific literature. These diverse modalities provide complementary insights, highlighting the relationships between processing, structure, property, and performance that are fundamental to materials science~\cite{hashemi2022computational}.

Historically, materials discovery has largely relied on structured data representations such as chemical formulae~\cite{kang2023multi,moro2025multimodal,wu2025versatile}, crystal structures~\cite{joshi2025allatom,ye2025materials,chenebuah2024deep,jiao2024space}, and graphs that represent local atomic environments and lattice connectivity~\cite{xie2018crystal,choudhary2021atomistic,deng2023chgnet}. These representations have been highly effective for training models that map structural features to target properties~\cite{ye2024cdvae,luo2024deep,khajeh2025materials}. However, these advances do not resolve the question of novelty, because material behaviours in real-world applications often depend on processing accessibility, microstructural realisation, and deployment constraints that are only partially captured by idealised structural representations.

Generative AI has further extended materials discovery by moving from property prediction to material candidate generation. Recent crystal diffusion models and related generative models can propose diverse and stable structures~\cite{lambard2023generation,altoyuri2024plastic}, while large language models (LLMs) can interpret textual descriptions of synthesis procedures and materials properties~\cite{jiang2025applications,schilling2025text}. However, generating novel materials requires more than crystal structure. 

Multimodal learning offers a route to address this limitation by allowing heterogeneous data sources to inform and constrain one another within a shared modelling framework~\cite{liu2025balancing,zhang2025artificial,liu2025towards}. Such integration can help preserve context, cross-scale relationships, and synthesis dependencies that are typically lost in single-modality models~\cite{jablonka2023ecosystem,omee2024structure}. It also lays a foundation for more credible inverse design, where generated candidates are evaluated not only for structural validity but also for their compatibility with broader scientific and experimental contexts~\cite {li2025probing}.

This Perspective explores whether generative and multimodal AI can support the design of novel materials. We first introduce a materials property hierarchy that distinguishes intrinsic properties from extrinsic performance, and use it to clarify the broader deployment constraints that shape materials discovery. Building on this framework, we propose a three-level taxonomy of structural, physical, and deployment novelty, highlighting that each level demands a different combination of evidence and a different role for multimodal AI. We then review multimodal materials data, recent advances in multimodal representation learning, knowledge integration, and generative AI, as well as current benchmark practices and their challenges. Finally, we identify four opportunities: multimodal data construction, process-aware modelling, feasibility-first generation and deployment-aware benchmarking. These opportunities outline a path from structural novelty towards deployment novelty. We conclude with recommendations for the materials and AI communities on the standards, infrastructure and reporting practices needed to strengthen the evidence base for novelty claims.

\begin{figure}[t]
\centering
\includegraphics[width=0.7\linewidth]{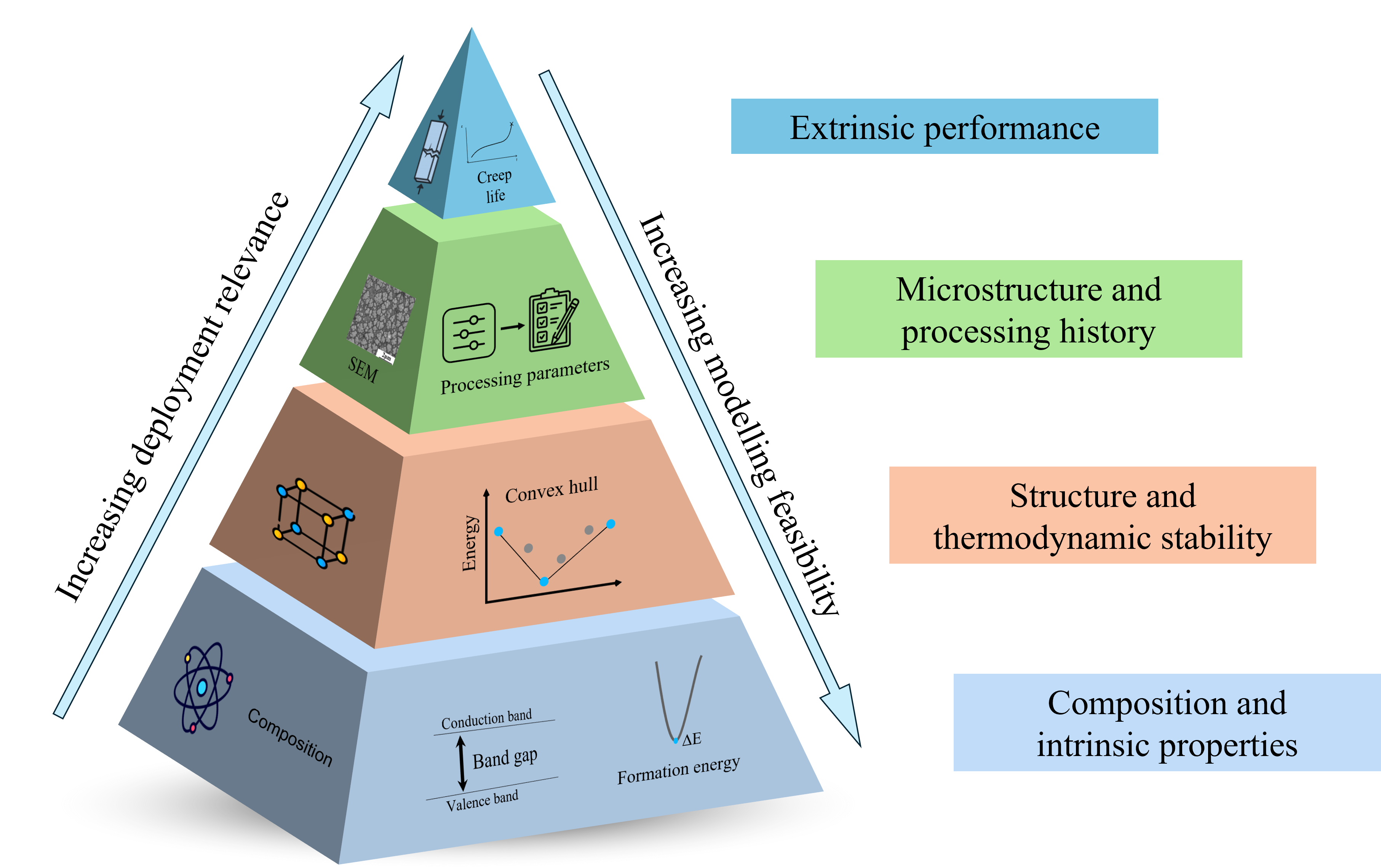}
\caption{\myrev{Hierarchical pyramid of materials property, from composition-determined intrinsic properties to processing-dependent extrinsic performance.}}
\label{fig:pyramid}
\end{figure}

% We focus on the joint promise of multimodal AI and generative AI for materials prediction and design: multimodal learning to unify fragmented information and constraints, and generative modelling to navigate design spaces more directly than screening alone. We summarise progress, emphasising what these models learn, what they still miss, and how they can be evaluated in ways that reflect end-to-end design outcomes. We also briefly highlight why this framing aligns naturally with deployment-oriented discovery workflows, including closed-loop experimentation and self-driving laboratories, where models must support decisions under practical constraints rather than operate as stand-alone predictors.

%-----------%
\section{From materials properties to novelty}
\label{sec:sec2-property}

Materials novelty cannot be defined without first specifying the behaviour a candidate is expected to exhibit. Intrinsic and extrinsic properties play different roles, while deployment constraints narrow the range of practically meaningful behaviours. This framing motivates the development of a taxonomy of novelty and evaluation criteria in this section.

\subsection{Materials property hierarchy}
\label{sec:property-hierarchy}

In many AI studies, materials are evaluated based on predicted or generated properties, which differ in their physical origins, observability, and relevance to deployment. Therefore, we distinguish between intrinsic and extrinsic properties within a hierarchical view of materials information, as illustrated in Fig. \ref{fig:pyramid}. At the lower levels are atomic-scale descriptors and intrinsic properties, which are closely tied to composition and idealised crystal structure. Higher levels contain microstructure, processing history, and engineering performance. As one moves up this hierarchy, prediction typically becomes less tractable, but also more relevant to real-world use. 

Intrinsic properties are primarily determined by the ground-state chemistry and atomic arrangement. When composition and phase are fixed, these properties are often relatively insensitive to processing history or microstructural variation. This is why crystal structure models and density functional theory (DFT) have been effective for predicting many electronic and thermodynamic quantities~\cite{jain2013commentary,kirklin2015open,saal2013materials,choudhary2020joint}. 

Extrinsic properties, by contrast, determine the engineering performance under practical conditions. Fracture toughness, fatigue life, and many functional responses depend not only on composition and crystal structure, but also on defects, interfaces, grain structure, phase morphology, and processing history.
These properties are path-dependent. Two materials with identical compositions and crystal structures can exhibit different mechanical or functional behaviours if they are processed differently. This path dependence highlights the limitations of unimodal or few-modality models that rely primarily on crystal structure, and it motivates the use of multimodal data that capture the full processing-structure-property chain.

This hierarchy is not proposed for a rigid boundary interpretation. In structural metallic alloys, the distinction is relatively clear. Intrinsic properties provide the thermodynamic foundation, while extrinsic performance is largely governed by microstructure and processing. However, in functional materials, intrinsic and extrinsic properties are often more strongly coupled. For example, in thermoelectrics, piezoelectrics, and battery electrode materials, key properties depend on both the electronic structure and microstructural features. Nevertheless, the hierarchy remains useful because it clarifies the organisation of material properties and behaviours and their relationship to materials data.

\subsection{Deployment constraints}

When the design objective shifts from intrinsic properties to extrinsic performance, deployment constraints emerge, including thermodynamic stability, processing accessibility, microstructural realisation, and practical synthesis constraints.

Computational materials discovery has traditionally focused on identifying thermodynamically stable phases. However, many technologically important materials rely on metastable structures that emerge through controlled processing routes.
Their value arises not from occupying the global energy minimum, but from being accessible via specific synthesis routes and retaining useful properties in practice. 
% Distance from the convex hull provides a useful first check but remains a rough proxy for synthetic feasibility.
\myrev{Distance from the convex hull is a stability-related measure for synthetic feasibility commonly used in generative models for crystal structure. It provides a useful first check, as candidates far above the hull are unstable and therefore less likely to be synthesised. However, it remains a rough, indirect proxy because stability is only one of the factors influencing synthesisability.}
Experimentally important materials often lie in metastable regions of the energy landscape, whereas some predicted ground states remain difficult to realise~\cite{Antoniuk_2023}.

The importance of metastability is evident across multiple material domains. In structural steels, useful combinations of high strength and ductility arise from microstructures produced through controlled thermomechanical processing rather than from equilibrium phase selection alone. Similarly, amorphous silicon and certain thin-film semiconductors often derive their technological value from non-equilibrium states stabilised by vapour deposition or rapid quenching~\cite{Walsh_2017, Xie_2024, Zheng_2025}. In such cases, the relevant design target is not only a stable crystal structure, but also a process-accessible state with useful behaviour.

More broadly, material behaviour is also influenced by coupled processing-structure-property relationships. Microstructural features, such as phase distribution, grain morphology, and defect structures, strongly influence performance. 
% Materials with identical compositions and crystal structures can exhibit different behaviour depending on their processing history.
Practical deployment also imposes synthesis constraints, including elemental availability, regulatory requirements, and compositional tolerances~\cite{zeni2025generative,Yang_2024,Hayes_2018,Mannan_2026}. These factors often determine whether a computational proposal remains a theoretical possibility or becomes a real material.

\myrev{In this Perspective, deployment is considered primarily in terms of synthesisability and experimental realisability, where synthesisability refers to whether a material can be produced through a plausible processing route, and experimental realisability refers to whether the resulting sample reproduces the targeted composition, structure, and microstructure under measurable conditions.}

\subsection{Materials novelty taxonomy}

The materials property hierarchy and deployment constraints shape the meaning of novelty in materials discovery. Statistical distance in composition or latent space is a reasonable screening criterion for intrinsic properties, but it does not measure extrinsic performance. A proposed material may sit far from any known composition and still be unstable, inaccessible, or functionally unremarkable, while a near-identical composition may behave very differently if it is processed differently. Therefore, novelty is not a single criterion. Following the materials property hierarchy introduced in Sec.~\ref{sec:property-hierarchy}, we organise material novelty in a hierarchical framework.

The first level is structural novelty, where a material occupies a previously unexplored region of compositional, crystallographic, or phase space. The second is physical novelty, where it exhibits properties, phase behaviour, or microstructural characteristics not observed in existing systems under comparable conditions. The third is deployment novelty, where it delivers performance that existing materials cannot achieve under realistic synthesis, manufacturing, or operating constraints.

These levels are relevant to different design objectives. Structural novelty may be sufficient for exploring new chemical or structural space, while physical novelty is important for identifying new intrinsic behaviour or distinct phase responses. Deployment novelty is required for designing experimentally realisable and practically useful materials. 

\myadd{\subsection{Materials novelty evaluation}}

\myadd{The evaluation at each novelty level requires different forms of material evidence, and also differs in how novelty can be established in practice. Structural novelty is the most tractable and can be evaluated against composition and structure alone, for example, using the distance to the nearest known material in a defined descriptor or structural space~\cite{betala2025lematgenbench,baird2024matbench}, combined with a basic stability check such as distance from the convex hull~\cite{Antoniuk_2023,gruver2024fine}. 
The definition of ``known'' depends on the criterion used for comparison. Structure-matching tools, such as \texttt{pymatgen}'s StructureMatcher~\cite{ong2013python}, are commonly used, while recent Pointwise Distance Distributions~\cite{widdowson2026pointwise} offer an alternative, identifying near-duplicates by comparing crystals as unordered point sets under isometry. Such comparisons are also sensitive to how disorder and reference structures are represented, as reflected in the recent discussion about the novelty of the MatterGen-proposed Ta-Cr-O candidate~\cite{juelsholt2026continued}.

Physical novelty evaluation requires a criterion in property rather than in input space. Based on a trusted predictor or measurement, a candidate material is physically novel only if its behaviour falls outside the reference distribution of known materials under comparable conditions, not only outside the training set~\cite{omee2024structure,li2025probing}. Predicted composition and crystal structure may be distinct and yet not synthesisable, in which case the associated behaviour is never realised. A property predicted from an idealised structure remains only a candidate for physical novelty until that behaviour is demonstrated in a material that can be made in practice~\cite{Antoniuk_2023}. In structural steel, for example, behaviour is determined jointly by the phases present, the microstructure produced by thermomechanical processing, and the resulting properties under relevant conditions, such as the combination of strength and ductility. A new steel composition occupying an unexplored region of the alloying space is only structurally novel. The same steel is physically novel only if a realisable, processed form exhibits a strength-ductility combination outside the range achieved by existing steels under comparable conditions, rather than a combination inferred from an idealised structure alone.

Deployment novelty sets a higher bar because its decisive properties are often extrinsic, path-dependent, and time-dependent. Evaluating these properties may require longer and more complex tests, making them difficult to incorporate into iterative model development. Surrogate proxies and early-stage indicators can support evaluation, but deployment novelty ultimately requires a demonstrable advantage over existing materials under realistic synthesis and operating conditions. For structural applications in particular, this advantage must hold not only in the as-synthesised or as-processed state but also throughout the service life, since properties can degrade through in-service exposure, including oxidation, fatigue, and corrosion. Therefore, deployment novelty extends to retained functionality under service-relevant ageing, rather than initial performance alone. This emphasis on retained functionality also connects to broader discussions of the distinction between a compound and a material, in which functionality and potential utility are central~\cite{cheetham2024artificial}. Establishing physical and deployment novelty requires information beyond composition and structure, including processing history, microstructure, synthesis records and measured performance. Since this information is distributed across different data types, multimodal data and AI workflows become essential to developing AI models for novel materials design.
}

\section{Multimodal materials data}
\label{sec:multimodal_data}

The hierarchy in Sec.~\ref{sec:sec2-property} shows that no single data type can fully characterise a material. Composition defines the chemistry, but processing, microstructure and characterisation together determine how that chemistry translates into observable behaviour. This section organises materials data into these four categories and introduces how they are currently represented, extracted and distributed.

\begin{table*}[t]
\centering
\scriptsize
\caption{Multimodal representations of materials data: A metallic alloy example}
\label{tab:representations}
\setlength{\tabcolsep}{3pt}
\renewcommand{\arraystretch}{1.2}
\begin{tabular}{l l l }
\toprule\toprule

\textbf{Data category} & \textbf{Data records} 
& \textbf{AI-ready encodings} \\
\midrule

\textbf{Chemical composition}
& \makecell[l]{Concentrations of each elemental constituent; \\ Formula and stoichiometry;\\ Precursor purity, ratios and composition ranges}
& \makecell[l]{Element-fraction vectors; Descriptor vectors;\\ Element tokens with learned embeddings;\\ Serialised composition text} \\
\midrule
\textbf{Microstructure}
& \makecell[l]{Microscopy images; \\ Chemical information and spectroscopy data; \\ Orientation and phase maps; \\Segmentations and region annotations}
& \makecell[l]{Patch or image embeddings; \\ Hyperspectral images and embeddings; \\ Multi-scale texture and morphology features;\\Region-level attributes; \\ Adjacency graphs between regions and/or phases} \\
\midrule
\textbf{Processing}
& \makecell[l]{Manufacturing process parameters;\\ Temperature, pressure, and time logs; \\ In situ sensor streams}
& \makecell[l]{Step sequences or action graphs; \\ Parameter vectors with protocol tags;\\ Time-series embeddings with event markers} \\
\midrule
\textbf{Testing and characterisation}
& \makecell[l]{Measured properties and curves; \\ Numerical mechanical property data;\\ Imaging from broken specimens; \\Instrument settings and analysis metadata}
& \makecell[l]{Curve and spectrum embeddings; \\ Peak and shape descriptors;\\ Metadata-aware features; \\ Fidelity tags for simulation vs experiment} \\

\bottomrule\bottomrule
\end{tabular}
\end{table*}

\subsection{Data modalities and representations}
\label{sec:materials_representations}

Materials data are inherently multimodal and can be broadly organised into four categories based on their data sources: \textbf{chemical composition, microstructure, processing, and testing and characterisation}~\cite{yu2025automated}. These categories are consistent with the broader materials property hierarchy. Each category may contain multiple modalities, and a single modality may contribute to several categories. Table~\ref{tab:representations} presents multimodal representations of metallic alloys. 

Chemical composition captures the elemental identities and proportions of a material and therefore constrains phase stability and intrinsic behaviour~\cite{yu2025automated}. It is commonly represented as chemical formulae, atomic-fraction vectors, or human-designed descriptors based on elemental properties~\cite{meredig2014combinatorial,ward2016general,butler2018machine,wang2024materials}. Composition is one of the most widely used material descriptors because it is specified before synthesis and remains relatively consistent across datasets. It is typically reported in a structured numerical form but can also be extracted from scientific texts and experimental reports~\cite{kononova2019text,lee2025text}.

Microstructure captures geometric and crystallographic information beyond that captured by composition alone. At the atomic scale, structural information is commonly represented using crystal structures encoded in formats such as Crystallographic Information Files (CIF), as atomic graphs~\cite{xie2018crystal,choudhary2021atomistic}, \myadd{or as 3D point clouds~\cite{teng2025atomic}}.
At larger length scales, microstructure comprises phase distributions, grain sizes, and crystallographic alignment, which are measured using microscopy and diffraction techniques and represented as images, voxel grids, or spatial fields~\cite{wissel2025stem,wanni2024machine,ling2017building}.

Processing information records the synthesis routes used to produce the material, which determine phase transformation kinetics and the emergence of non-equilibrium states. This information varies significantly across material domains. Processing information is represented as synthesis text, equipment-generated time series, or structured provenance records that capture dependencies between fabrication steps~\cite{lee2025text,tsuruta2025matprov,kim2025towards,zhang2025advancing,deshmukh2024effect,lederbauer2025lematsynth}.

Testing and characterisation observations record the material behaviour under external stimuli and operating conditions. These observations include instrument- or simulator-level outputs, such as mechanical, electrical, magnetic, or spectroscopic responses~\cite{salgado2023automated,oviedo2019fast}. As they connect material state to measured function, they are essential for validating whether a proposed material meets its intended performance requirements~\cite{paulus2023towards,statt2023materials}. 

\begin{figure}[t]
\centering
\begin{subfigure}{0.47\linewidth}
    \centering
    \includegraphics[width=\linewidth]{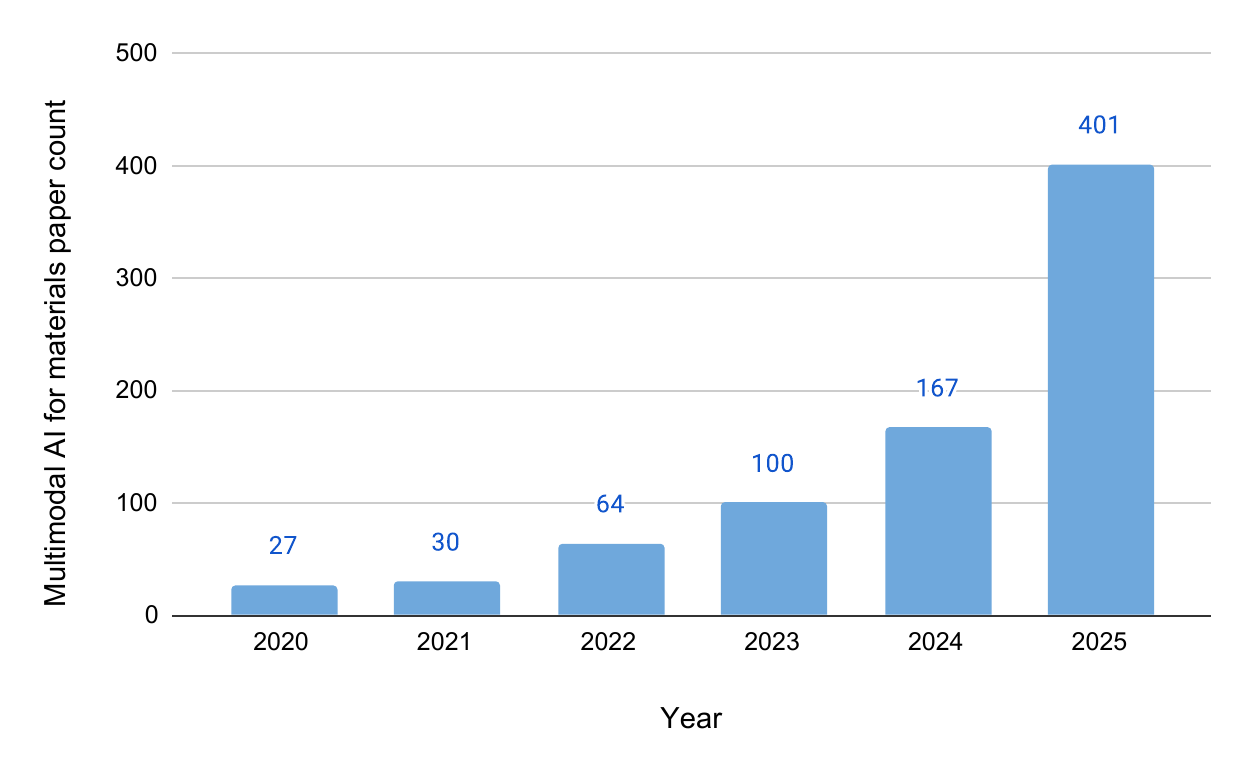}
    \caption{}
    \label{fig:category-trend}
\end{subfigure}
\hfill
\begin{subfigure}{0.47\linewidth}
    \centering
    \includegraphics[width=\linewidth]{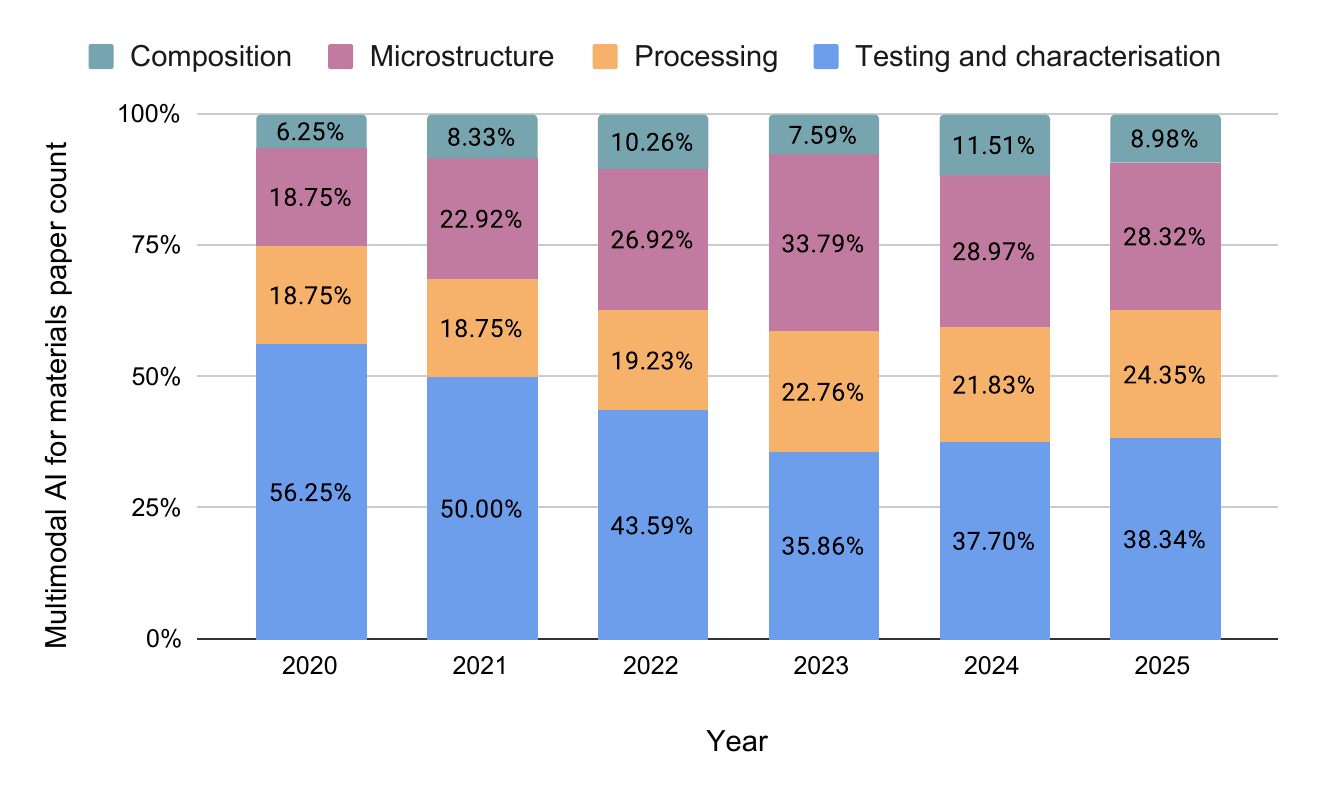}
    \caption{}
    \label{fig:category-details}
\end{subfigure}
\caption{\myrev{Multimodal AI for materials publications by data category from 2020 to 2025. (a) The annual number of publications that match the multimodal AI for materials query and involve at least one of four materials data categories. (b) Relative proportion of the data categories each year, where publications spanning multiple categories are counted in each.
Publication data were retrieved from the Elsevier Scopus database on 24 June 2026. See Supplementary Information Section 1 for details of the trend analysis.}}
\label{fig:category}
\end{figure}

Figure~\ref{fig:category}a shows the annual growth in publications that incorporate these four categories. Their representation remains uneven, as shown in Fig. \ref{fig:category}b. Testing and characterisation dominate, while microstructure has expanded in recent years. Composition remains the smallest component throughout the period, despite appearing in most multimodal studies, which reflects that it is often used as baseline information combined with other data types rather than analysed as a primary data input.

\myadd{\subsection{Multimodal data extraction}}

\myadd{Across all four categories, materials data are rarely available in a structured, AI-ready form at the point of generation. Scientific literature remains one of the most important sources for data extraction, but relevant information is often dispersed across text, tables and figures, making acquisition and curation a major bottleneck. Literature-mining pipelines, such as ChemDataExtractor~\cite{swain2016chemdataextractor,mavracic2021chemdataextractor}, use named entity recognition and rule-based parsing to extract chemical entities and properties into structured databases. 
ChemDataExtractor remains widely used for its high precision and reproducibility, but its reliance on hand-crafted rules requires substantial domain expertise when extending it to new material classes.
More recently, LLMs have helped to reduce this manual effort and improve extraction flexibility, including conversational prompting to extract and self-verify properties~\cite{polak2024extracting} and fine-tuned LLMs for structured entity and relation extraction~\cite{dagdelen2024structured}. Recent approaches are pushing extraction towards multimodal and larger-scale settings. LeMat-Synth~\cite{lederbauer2025lematsynth} combines LLMs and vision-language models to extract synthesis procedures and performance data from both text and figures. MatSKRAFT~\cite{hira2025matskraft} uses constraint-driven graph neural networks (GNNs) to assemble large materials knowledge bases from scientific tables. These approaches determine which modalities can be collected at scale and, consequently, shape the data foundation available for AI model development.}

\subsection{Coverage, imbalance, and heterogeneity}
\label{sec:challenges_multimodal_data}
Despite these advances, multimodal materials data analysis faces several challenges, including limited coverage, modality imbalance, missing negative results, and heterogeneity across laboratories and measurement pipelines~\cite{Li2026, Debnath2021}. These challenges constrain both the relationships that multimodal models can learn and their ability to generalise reliably beyond the training data.

Materials datasets are often scarce and concentrated in specific compositional or processing subspaces because synthesis and characterisation are expensive and time-consuming~\cite{wu2025versatile, Li2026, Reeves-McLaren2026, Berry2025, Wang2020, Chang2022, Xu2023}. The imbalance exists within the synthesis and characterisation data. For example, microstructural imaging and rapid measurements, such as hardness, are widely reported, whereas detailed processing metadata, failed synthesis records, and long-duration performance measurements are much less available~\cite{Reeves-McLaren2026, che2023deep, Wong2025, Gianola2023}. In metallurgy and alloy development, long-term tests, such as creep, are particularly difficult to collect at scale because a single experiment may require 10$^{3}$-10$^{4}$ hours~\cite{Mouritz2012}.
Consequently, available datasets usually capture only part of the processing-structure-property relationships, and are often dominated by the most accessible data types, which may not be the most informative for deployment-relevant behaviour.

\myadd{This imbalance is also visible in widely used resources. Large open databases, such as the Materials Project~\cite{jain2013commentary}, OQMD~\cite{kirklin2015open,saal2013materials} and JARVIS~\cite{choudhary2020joint}, provide large-scale, AI-ready data on compositions, idealised crystal structures and DFT-derived intrinsic properties, but offer limited coverage of processing and characterisation information~\cite{horton2025accelerated}. Microstructures are expanding rapidly but remain fragmented across smaller repository depositions and literature-mined collections~\cite{liu2026multimodal}. More comprehensive multimodal and processing-aware resources may exist on commercial materials information platforms, but they remain proprietary and largely inaccessible to open model development.}

A related challenge is the systematic absence of negative or null results, namely ``dark data''. Failed syntheses, unsuccessful processing routes, and compositions that do not produce the intended phase or property are rarely reported, even though they are important for constraining design space and defining feasibility boundaries~\cite{Boyce2023,maloney2023negative, Kozlowski2022}. 

Materials datasets are also highly heterogeneous across laboratories, instruments, and analysis pipelines. Experimental outputs depend strongly on sample preparation, instrument calibration, and downstream processing, while quality-control and reporting practices vary substantially between laboratories~\cite{Himanen2019}. Without consistent metadata standards and detailed reporting of experimental conditions, it becomes difficult to compare results across sources or to distinguish physically meaningful variation from artefacts introduced during acquisition and analysis.

Taken together, these challenges show that multimodal materials datasets are not only limited in quantity but also uneven in coverage, selective in reporting, and variable in quality across sources.

\section{Multimodal learning and generative AI for materials}
\label{sec:multimodal_gen_AI}

\begin{figure}[t]
 \centering
\includegraphics[width=0.7\linewidth]{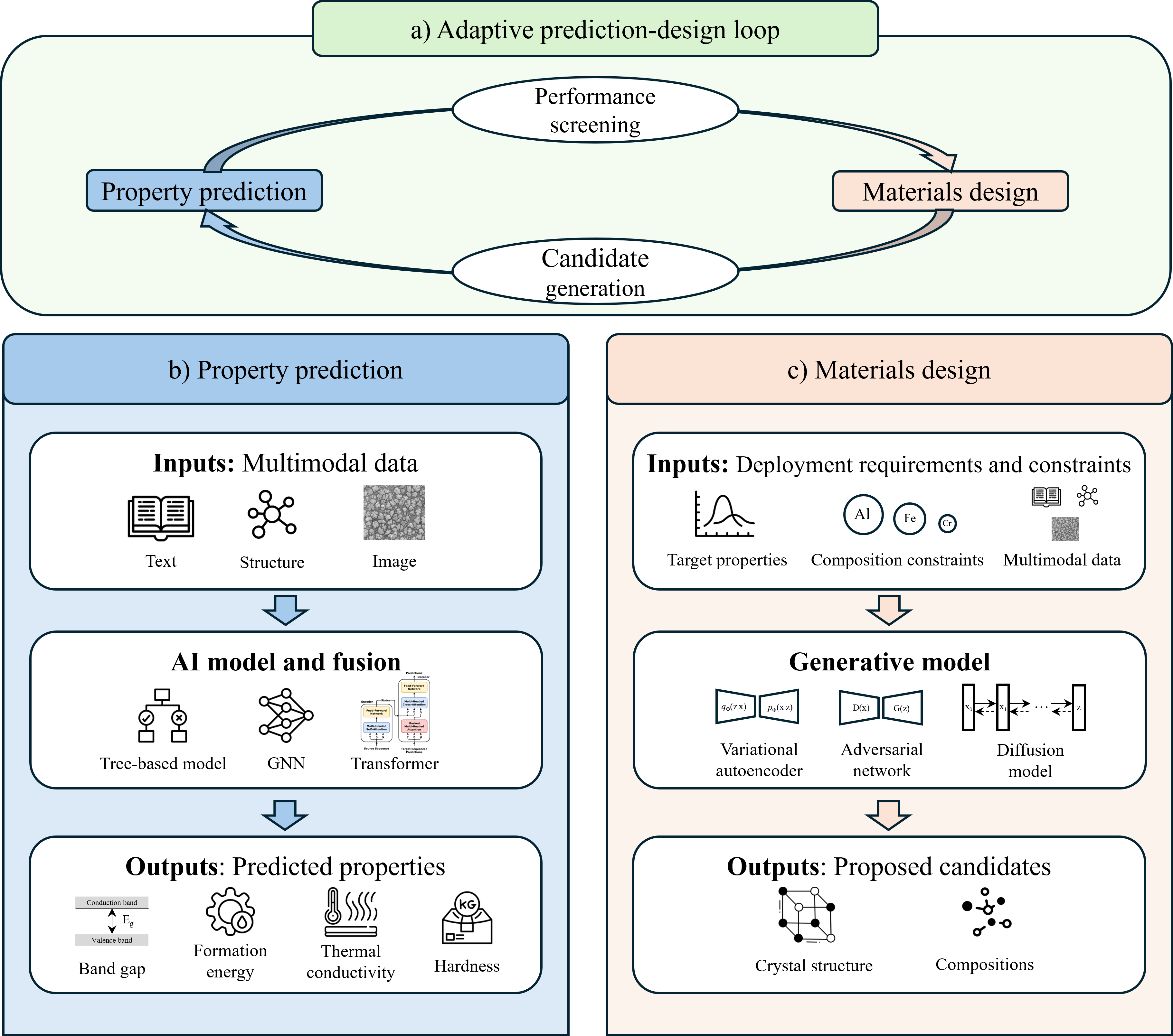}
\caption{Generative and multimodal AI for materials discovery, including adaptive prediction-design loop (a), property prediction (b), and materials design (c).}
\label{fig:loop}
\end{figure}

The prior sections demonstrate that designing novel materials requires broader evidence from the materials property hierarchy, and that this evidence is distributed across heterogeneous data sources. Current AI for materials research can be organised around two coupled tasks: property prediction and materials design. As shown in Fig. \ref{fig:loop}, these tasks form an adaptive prediction-design loop in which candidate generation and performance screening iteratively inform each other. Within this loop, multimodal learning provides the foundation for extracting, aligning and fusing information across heterogeneous materials data, while generative AI can use these representations to propose new compositions, structures, or processing routes. Knowledge graphs and other knowledge-integration frameworks can effectively learn fragmented relational knowledge from the literature and experiments to support prediction and design. The following reviews recent progress in multimodal representation learning and fusion, knowledge integration, and generative AI for materials prediction and design.

\subsection{Multimodal representation learning for materials prediction}
\label{sec:multimodal_representation_learning}

Multimodal learning begins by encoding heterogeneous materials data into AI-ready representations, and different data types usually require different encoders.

Chemical composition can be represented either through handcrafted elemental descriptors used with conventional machine learning models~\cite{goodall2020predicting} or through neural networks that learn directly from formula tokens and fractional amounts~\cite{wang2021compositionally,jha2018elemnet}. 
Crystal structures are commonly encoded as periodic graphs for GNNs that aggregate local coordination environments into structure-aware representations~\cite{xie2018crystal, choudhary2021atomistic, deng2023chgnet}. Because composition and crystal structure are generally more accessible and more consistently curated than processing or characterisation data, these methods are widely used across many materials domains.

Processing information, such as synthesis protocols or thermomechanical histories, can be represented as ordered sequences or provenance graphs extracted from laboratory records and scientific literature~\cite{deshmukh2024effect,kim2017materials,wang2022dataset}.
Microstructural observations are commonly encoded using vision models, including convolutional neural networks and vision transformers, to extract quantitative representations from microscopy images and orientation maps~\cite{wissel2025stem,wanni2024machine,ling2017building,che2023deep}.
One-dimensional experimental signals, such as X-ray diffraction patterns, can be processed with convolutional or attention-based sequence models~\cite{paulus2023towards,statt2023materials,vecsei2019neural,simonnet2024phase}.
Textual records can be analysed using domain-specific language models to extract entities, relationships, and synthesis-relevant constraints~\cite{gupta2022matscibert, shetty2023general}.

Recent work has also explored LLMs as a more unified interface for heterogeneous materials data. In these approaches, non-text inputs are converted into structured textual descriptions that can be processed within a shared language representation. This can simplify multimodal pipelines, although it may also sacrifice numerical precision or spatial detail~\cite{niyongabo2025llm}.

Once modality-specific representations are learned, the next challenge is to align and fuse them into a shared representation that preserves complementary information across modalities. Cross-modal alignment aims to ensure that embeddings from different modalities correspond to the same material state, often using contrastive learning to pull matched pairs together and push mismatched pairs apart. In practice, this step is difficult, as modalities can differ in completeness, resolution, invariance, and processing dependency. Textual descriptions can be incomplete or ambiguous, and experimental measurements may correspond to different processing states.
Fusion then combines aligned representations to make predictions. Simple concatenation remains a common baseline, while more advanced methods use attention or gating mechanisms to weight modalities based on their information content and reliability.

Multi-fidelity learning further extends this framework by integrating computational and experimental materials data, which often provide correlated but systematically biased views of material behaviour~\cite{jiang2026multi}. This is especially relevant in materials research, where broad simulation coverage coexists with smaller but higher-fidelity experimental datasets.

Overall, multimodal representation learning provides the predictive foundation for linking composition, structure, processing, and observation within a common modelling framework. By integrating complementary views of material state and behaviour, these methods can yield more robust predictions than unimodal models alone, and the multimodal representations subsequently serve as the conditioning signal that enables generative models to propose candidates compatible with these contexts (Sec.~\ref{sec:genAI}).

\subsection{Knowledge integration via materials knowledge graphs}
\label{sec:knowledge_integration}

While multimodal representation learning focuses on encoding materials data into shared representations, scientific knowledge is often dispersed across literature, databases, and experimental records, rather than appearing as aligned training pairs. This has motivated growing interest in materials knowledge graphs, which provide a structured way to integrate materials data with broader scientific context~\cite{hira2025matskraft,matkg,matkg2,matkg3,matkg4,matkg5}.
Materials knowledge graphs organise connected entities such as materials, properties, processes, synthesis and characterisation methods, phase labels, applications, and scientific documents within a common relational framework~\cite{matkg}. This graph structure provides a way to represent a scientific context that is difficult to capture from composition or structure alone.

Materials knowledge graphs can contribute at three levels. First, their relational structure captures connections between materials, properties, processes, and other scientific entities that are not available from internal composition or crystal structure alone, thereby supporting higher-order reasoning and more grounded property prediction~\cite{matkg6}. Second, knowledge graphs can serve as an external knowledge source for materials language models, allowing updated information to be retrieved at inference time rather than relying entirely on static training data~\cite{jiang2025applications,matlm,fu2023material,matlm4,kara,rag,selfcheckgpt,llm_hallucination}. Third, by linking structured data, text, and experimental records within a shared framework, they provide a practical route for integrating fragmented multimodal scientific knowledge~\cite{multimodal_kg,multilingual_kg}.

Overall, materials knowledge graphs support relational reasoning, knowledge grounding, and heterogeneous data integration, connecting composition and structure to the synthesis and characterisation context that often determines whether a material is scientifically meaningful and practically relevant.

\subsection{Generative AI for multimodal materials design}
\label{sec:genAI}

\myadd{Multimodal representations and knowledge structures provide the basis for generative modelling. The encoders and fusion strategies introduced in Sec.~\ref{sec:multimodal_representation_learning} for property prediction return here in a different role: they define the conditioning space used to propose new candidates. This allows generated outputs to be constrained not only by chemical validity but also by the processing, microstructural, and performance contexts associated with the design target.
The following reviews recent generative architectures that use multimodal conditioning to move from prediction to candidate generation, and discusses the constraints used in the generation process.
}

Generative AI is reshaping materials discovery by proposing new compositions, structures, and processing-relevant candidates rather than only predicting the properties of known materials~\cite{cheng2026artificial}. Generative models support both unconditional and conditional generations. Unconditional models explore the space of chemically valid materials without predefined targets. Conditional models guide generation towards desired properties or synthesis constraints, such as band gap, crystal symmetry, or synthesis-related requirements.

Several generative families have been explored for these tasks~\cite{zeni2025generative,xie2022crystal, alverson2024crystens}. Variational autoencoders learn continuous latent spaces that support interpolation and optimisation. Generative adversarial networks enable fast sampling but often suffer from instability. Diffusion models iteratively refine noisy inputs into structured outputs and have become prominent for crystal generation. 
LLMs provide a complementary interface by treating materials as tokenised sequences, such as formulae, crystal descriptions, or synthesis protocols, allowing generation to be conditioned on textual or structured prompts~\cite{mohanty2026crystext}. 
\myadd{LLMs for materials generation range from generating chemical formulae~\cite{fu2023material} to crystal structures as text~\cite{gruver2024fine}. However, their use also raises concerns about data leakage from large-scale pretraining and hallucinations that can introduce invalid structures, thereby motivating fine-tuning, validity checks, and post-generation filtering.}
These methods aim to move from passive prediction to active proposal of candidate materials.

Recent progress has been driven particularly by conditional generation. In materials design, useful conditions often combine hard constraints, such as stoichiometry, symmetry, or lattice consistency, with soft objectives, such as stability proxies, target property ranges, or dopability heuristics. 
Diffusion-based frameworks have proven flexible in this setting because constraints can be incorporated through training, guided sampling, or constrained decoding~\cite{zeni2025generative, jiao2024space, okabe2026scigen}. 
Related methods also treat generation as a constrained completion problem, for example, by reconstructing missing structural information from partial crystallographic or experimental evidence~\cite{reents2026inpainting}. 
\myadd{A further route is external-objective guidance, in which a pretrained generator is steered at sampling time using a surrogate model or reward, without retraining the conditional model. This idea is well established in image generation~\cite{chung2023diffusion} and is being adapted to materials generation through reinforcement-learning guidance using stability or target-property rewards~\cite{park2025guiding}.}

More broadly, these developments shift part of the upstream feasibility problem into the generator rather than relying entirely on post-hoc filtering.
\myadd{For example, several recent models embed symmetry directly into the generative process. CrystalFormer~\cite{cao2025space} generates space-group-controlled crystal structures by predicting Wyckoff positions, chemical elements, and fractional coordinates. SymmCD~\cite{levy2024symmcd} models the asymmetric unit with the symmetry transformations for each atom in the diffusion process. LEGO-xtal~\cite{ridwan2026ai} uses space-group and Wyckoff-based representations with local-environment targeting to build complex structures from only limited known examples.}

This shift toward generation for feasibility is important because inverse design is inherently multimodal. A target performance profile is not translated into a deployable material solely through structure, but through the coupled selection of composition, processing route, microstructure, and expected response. In this setting, generative models need to do more than sample valid crystal structures. They also need to incorporate the conditions under which those structures can be realised and whether they remain compatible with downstream synthesis and performance constraints. Therefore, multimodal representations and knowledge integration provide the context needed to connect generated candidates to feasible material states rather than to abstract points in the design space.

\myadd{Generative models are increasingly applied beyond structure proposal to synthesis itself. Diffusion models such as DiffSyn~\cite{pan2026diffsyn} generate synthesis routes, including compositions and synthesis conditions, conditioned on a target structure. LLMs have been used to propose inorganic synthesis conditions and precursors~\cite{prein2025language}, and reinforcement learning with experimental feedback has been used to align reasoning traces with experimental outcomes~\cite{pan2026synreason}.}

These developments also change how generative models should be evaluated. Raw validity, diversity, or latent-space novelty are not sufficient if generated candidates fail under stability screening, violate symmetry or synthesis constraints, or cannot be connected to realistic processing routes. Therefore, emerging benchmarks increasingly evaluate trade-offs among validity, novelty, diversity, and constraint satisfaction under more realistic discovery settings~\cite{betala2025lematgenbench}. This motivates the discussion in Sec.~\ref{sec:benchmark}, where we examine benchmarking criteria that prioritise deployability over single-metric optimisation.

\section{Evaluation and benchmarking}
\label{sec:benchmark}

Benchmarks make models testable, but only when their reference data reflect the real-world discovery setting they are designed to evaluate. In AI for materials, this alignment remains incomplete. This section shows that many tasks rely on computational labels, that generative evaluation often depends on proxy metrics, and that experimental utility is rarely measured directly.

\subsection{Evaluation and benchmarking landscape}
Evaluation of materials AI models typically combines task definition, data splitting, and a set of performance metrics. Property prediction tasks are often framed as regression or classification, with error-based, correlation-based, or threshold-based metrics used for comparison. Most studies still rely on random held-out splits or cross-validation, but these protocols can overestimate practical utility when test conditions differ from those seen during training. More recent evaluations use out-of-distribution settings that better reflect deployment conditions and evaluate the novelty of model predictions~\cite{jablonka2023ecosystem,omee2024structure,li2025probing,witman2025matfold}. In alloy systems, this can be framed in physically meaningful terms, such as whether a model trained on binary and ternary spaces can generalise to quaternary or higher-order alloys, or whether it can transfer across processing routes for nominally similar compositions. Such settings are more demanding than random splits, but they more closely reflect the challenges of real-world discovery.

Generative modelling requires a broader evaluation framework because outputs are evaluated both as samples from a learned distribution and as candidates for downstream screening. Common metrics include validity, novelty, uniqueness, and coverage, and recent toolkits adopt more consistent, leakage-resistant protocols, including time-aware splits for crystal datasets~\cite{baird2024matbench}. Deployment-oriented evaluation extends these criteria by including feasibility checks, such as chemical validity, structural consistency, and stability proxies under a chosen oracle, and by asking whether generated candidates satisfy the intended property or structural constraints rather than collapsing to easy-to-generate modes~\cite{de2025generative}. For conditional generation, useful metrics also include the success rate under target constraints and the best-of-$N$ utility under a fixed screening budget, since practical workflows rank and filter batches rather than accept individual samples in isolation.

\begin{table*}[t]
\centering
\tiny
\caption{Representative benchmarks for AI in materials research, grouped by task type and summarised by material domain, reference signal, evidence categories and data modalities. \textit{Comp.}, \textit{Microstruct.}, \textit{Proc.} and \textit{Test. \& character.} denote the primary data categories. \textit{Text}, \textit{Graph}, \textit{Image}, and \textit{Spectra} denote the data modalities used to instantiate these data categories. The asterisk (*) indicates that multi-fidelity data are present, but paired settings are not provided, so multi-fidelity learning could be constructed when matched entries exist.
}
\label{tab:benchmarks}
\setlength{\tabcolsep}{3pt}
\renewcommand{\arraystretch}{1.2}
\begin{tabular}{l c l l ccccc | cccc}
\toprule\toprule

\multirow{2}{*}{\textbf{Benchmark}} 
& \multirow{2}{*}{\textbf{Year}} 
& \multirow{2}{*}{\textbf{Materials domain}} 
& \multirow{2}{*}{\textbf{Reference signal}}
& \multicolumn{5}{c}{\textbf{Primary evidence categories}}
& \multicolumn{4}{c}{\textbf{Primary data modalities}} \\

& & & 
& \multicolumn{1}{c}{\textbf{Comp.}} 
& \multicolumn{1}{c}{\textbf{Microstruct.}} 
& \multicolumn{1}{c}{\textbf{Proc.}} 
& \multicolumn{1}{c}{\textbf{Test. \& character.}} 
& \multicolumn{1}{c}{\textbf{Multi-fidelity}} 
& \multicolumn{1}{c}{\textbf{Text}} 
& \multicolumn{1}{c}{\textbf{Graph}} 
& \multicolumn{1}{c}{\textbf{Image}} 
& \multicolumn{1}{c}{\textbf{Spectra}} \\

\midrule
\multicolumn{13}{c}{\textbf{A. Property prediction}} \\
\midrule

\makecell[l]{Matbench Discovery~\cite{riebesell2025framework}}
& 2025
& Crystals
& DFT
& \cmark & \cmark &  &  & 
&  & \cmark &  & \\

TextEdge~\cite{niyongabo2025llm}
& 2025
& Crystals
& DFT
& \cmark & \cmark &  &  &
& \cmark &  &  & \\

JARVIS-Leaderboard~\cite{choudhary2024jarvis}
& 2024
& Broad materials
& \makecell[l]{DFT \\ Experimental}
& \cmark & \cmark &  & \cmark & \cmark*
& \cmark & \cmark & \cmark & \cmark \\

Matbench~\cite{dunn2020benchmarking}
& 2020
& Broad materials
& \makecell[l]{DFT \\ Experimental}
& \cmark & \cmark &  &  & \cmark*
&  & \cmark &  & \\

\midrule
\multicolumn{13}{c}{\textbf{B. Generative modelling}} \\
\midrule

LeMat-GenBench~\cite{betala2025lematgenbench}
& 2025
& Crystals
& Stability proxies
& \cmark & \cmark &  &  &
&  & \cmark &  & \\

MGB~\cite{yan2025mgb}
& 2025
& Crystals, MOFs
& Proxy metrics
& \cmark & \cmark &  &  &
&  & \cmark &  & \\

matbench-genmetrics~\cite{baird2024matbench}
& 2024
& Crystals
& Proxy metrics on time splits
& \cmark & \cmark &  &  &
&  & \cmark &  & \\

\midrule
\multicolumn{13}{c}{\textbf{C. Multimodal interpretation}} \\
\midrule

MATRIX~\cite{mcgrath2026matrix}
& 2026
& Materials reasoning
& Prompts and rubrics
& \cmark & \cmark & \cmark & \cmark &
& \cmark &  & \cmark & \cmark \\

% ChemBench 
% & Scientific reasoning 
% & 2025 
% & Chemistry 
% & Curated QA 
% &  &  &  & \cmark & 
% & \cmark &  &  &  \\

MaCBench~\cite{alampara2025probing}
& 2025 
& Materials reasoning
& Tasks and solutions
& \cmark & \cmark & \cmark & \cmark &  
& \cmark &  & \cmark & \cmark \\

MatQnA~\cite{weng2025matqna}
& 2025
& Materials characterisation
& QA pairs from papers
&  &  &  & \cmark &
& \cmark &  & \cmark & \cmark \\

% \midrule
% Open Catalyst
% & Interatomic potentials
% & 2020--2022
% & Catalysis surfaces
% & DFT energies
% &  & \cmark &  &  &
% &  & \cmark &  & \\

\bottomrule\bottomrule
\end{tabular}
\end{table*}

Benchmarks package these evaluations into a shared protocol by defining tasks, data splits, evaluation metrics, and baseline methods, often with public leaderboards. Table~\ref{tab:benchmarks} summarises representative benchmarks by task type, materials domain, reference signals, primary input categories, and dominant modalities. Property-prediction benchmarks remain strongly shaped by crystalline materials with computational labels, where composition and crystal structure serve as the primary inputs and graph-based encodings are common. Suites such as Matbench~\cite{dunn2020benchmarking}, JARVIS-Leaderboard~\cite{choudhary2024jarvis}, and Matbench Discovery~\cite{riebesell2025framework} have improved standardisation, while TextEdge~\cite{niyongabo2025llm} begins to extend crystal property prediction towards text-based inputs. Large-scale community benchmarks also exist for interatomic potentials and catalysis, such as the Open Catalyst Project~\cite{chanussot2021open,tran2023open}. By contrast, generative benchmarks still rely heavily on proxy validation, particularly in inorganic crystal generation~\cite{betala2025lematgenbench,baird2024matbench,yan2025mgb}. Emerging multimodal and LLM-based benchmarks increasingly probe cross-modal reasoning and interpretation of experimental artefacts~\cite{mcgrath2026matrix,alampara2025probing,weng2025matqna}, but process data remain under-represented relative to composition, structure, and observation.

\subsection{Benchmarking challenges for deployability}
\label{benchmarking_challenge}

Despite this growing benchmark ecosystem, several challenges remain.
A fundamental challenge is the reliance on computational reference labels, such as DFT or related simulations. These benchmarks are valuable because they are scalable and internally consistent, but they only partially reflect experimental deployment. For example, for structural alloys, benchmark labels are often computed for idealised, defect-free unit cells at zero Kelvin, while real-world performance depends on finite-temperature phase stability, grain-boundary chemistry, dislocation structures, and thermomechanical history. Strong performance on computational benchmarks does not necessarily translate into laboratory utility and may even reward shortcuts specific to the computational label distribution. The evaluation of generative models often depends on computational oracles or surrogate predictors, which introduce systematic bias and only partially reflect experimental feasibility~\cite{perdew1985density,harvey2006accuracy,lejaeghere2014error}.

Out-of-distribution evaluation also remains difficult to define in a physically meaningful and standardised way, making it insufficient to evaluate the novelty of model predictions. Although guided by domain knowledge, some out-of-distribution splits can still leave substantial overlap between training and test data in representation space. For example, this can happen when held-out alloy families share structural motifs and chemistry with the training set, or when test points fall within high-density regions of the compositional space already covered by training data. Subsequently, reported out-of-distribution performance reflects interpolation rather than true extrapolation, inflating claims of transferability~\cite{li2025probing}.

Reporting practices remain inconsistent across benchmark studies. Computational cost, novelty-stability trade-offs, verification budgets, and success rates under constrained validation are not always reported in a comparable manner. Unified benchmark suites improve standardisation, but deployability will require evaluation protocols that connect accuracy and generative quality to experimentally relevant criteria, including feasibility, robustness, and performance under limited verification resources. 
\myadd{Task-specific benchmark suites provide useful precedents, e.g., CSPBench~\cite{wei2024cspbench}, which shows that model rankings can depend strongly on test-set construction and structural similarity metrics.}

The benchmarking challenges share a deeper common cause. Computational validation and experimental validity are not points on a single fidelity continuum but measures of fundamentally different things. \myadd{Computational labels typically describe an idealised model system, such as a defect-free unit cell evaluated at zero Kelvin under a chosen exchange-correlation approximation. Experimental measurements instead reflect the behaviour of fabricated materials, which depends on finite temperature, defects, and processing history. Although these factors can be simulated, they usually fall outside the idealised setup used to generate benchmark labels. Therefore, the resulting computational and experimental targets are not simply low- and high-fidelity versions of the same quantity.}
\myrev{Higher performance against computational benchmarks may not translate to experimental validity~\cite{cheetham2024artificial}}. A model validated against computational reference signals has been tested only on its ability to reproduce a specific theoretical approximation of material behaviour. This provides limited evidence that the learned representations will transfer to laboratory conditions, because systematic errors in computational references can be correlated across training and test distributions in ways that experimental measurements are not~\cite{perdew1985density,harvey2006accuracy,lejaeghere2014error}.

These challenges show that benchmark performance is not necessarily correlated with deployable utility, while also clarifying the direction of progress. As evaluation moves beyond random splits, single-metric accuracy, and proxy-only validation, more realistic criteria are needed to evaluate whether generative and multimodal models can support the discovery of novel materials.

\section{Future opportunities and recommendations}

\myrev{
% \subsection{Remaining challenges}
% \label{sec:challenges}

A consistent pattern emerges from the preceding sections. AI for materials research has made substantial progress on structural novelty, with generative models proposing compositionally and crystallographically diverse candidates at scale. However, progress slows markedly when moving up the hierarchy towards physical and deployment novelty. The remaining challenges are largely due to insufficient evidence grounding across the hierarchy. The evidence available to models is concentrated at the level of composition and idealised structure, whereas the evidence required to establish higher levels of novelty remains under-represented and weakly integrated.

The transition from structural to physical novelty is largely a data-coverage challenge with two key aspects. The first is the scarcity of processing and microstructural data: the modalities required to support physical novelty are far less available than compositional and structural data (Sec.~\ref{sec:challenges_multimodal_data}). This scarcity reflects the economics of experimental materials science, where high-throughput screening or computation is relatively accessible, while detailed characterisation and long-duration testing remain costly and slow. Even when such modalities are reported, they are often coarse or poorly aligned with the composition and structure, limiting their value for attributing or comparing physical behaviour. The second is the scarcity of negative data, where ``dark data'' remains largely unreported under current success-oriented publication incentives. Without these counterexamples, models are susceptible to learning the interior of known compositional space, rather than the boundaries of what is physically realisable~\cite{maloney2023negative}. Generative models trained on such data can therefore support structural novelty but cannot establish whether a candidate is physically distinct from known materials. 

The transition from physical to deployment novelty adds architectural and benchmarking challenges on top of these data challenges. Many multimodal models combine independently encoded modalities without learning the cross-modal interactions that govern deployment-relevant behaviour. Processing and microstructure, for example, are physically coupled: a given thermomechanical processing sequence produces a specific grain boundary chemistry, and that link is a causal, physical one. A model that encodes processing and microstructure separately and simply concatenates the resulting embeddings may not adequately capture the relationship between them. Benchmarking poses a parallel challenge, as current benchmarks largely rely on computational proxies rather than experimental measurements, providing limited evidence of transfer to laboratory synthesis, manufacturing, or in-service conditions (Sec.~\ref{benchmarking_challenge}).

A cross-cutting challenge underlying the three levels is a lack of standardisation. Data and metadata are reported in diverse, often incompatible formats, making it difficult to assemble consistently annotated multimodal datasets that span the full processing-structure-property chain. Without standardised schemas for consistent data reporting and an open infrastructure for data sharing, deployment novelty will remain poorly grounded, regardless of advances in AI models. Addressing these challenges requires community-wide infrastructure development.

\subsection{Four opportunities for evidence-grounded discovery}

The most immediate opportunity is to build standardised multimodal datasets that link each material to the context needed to interpret it. In such datasets, each material record should connect core data (composition, structure, microstructure, and properties) to the corresponding provenance metadata (e.g., synthesis route, processing conditions, measurement settings, units, uncertainties, and fidelity levels). This metadata should be captured and clearly reported at the source, because they are hard to recover afterwards, a key limitation of datasets assembled from the literature. To assemble such datasets, a standardised workflow is needed for generating and reporting experimental data, with shared schemas and reporting templates that keep modalities aligned, adapt across material classes and experiment types, and follow the FAIR principles~\cite{wilkinson2016fair}. Emerging closed-loop and self-driving laboratories, such as A-Lab~\cite{szymanski2023autonomous} and AlphaFlow~\cite{volk2023alphaflow}, provide an automated version of this workflow, acting not only as discovery platforms but also as generators of aligned multimodal data, including failed trials. Once assembled, these datasets can support multimodal data fusion and knowledge integration (Sec.~\ref{sec:multimodal_representation_learning} and Sec.~\ref{sec:knowledge_integration}).

Beyond constructing datasets, the next opportunity is to develop process-aware multimodal AI models that treat processing and manufacturing information as a primary modality rather than peripheral metadata. This can advance the model towards learning the full processing-structure-property relationship, rather than only mapping composition or crystal structure to a single target property. However, the processing modality is often scarce, coarse or inconsistently reported, requiring models to be robust to missing or heterogeneous inputs. Uncertainty-weighted fusion~\cite{han2022trusted}, for example, can down-weight unreliable inputs. Alongside statistical weighting, physical constraints can help compensate for missing data. Different modalities are often connected by underlying physical relationships, so a missing modality can sometimes be constrained or inferred using known physical laws. For example, thermodynamic phase rules can be used to partially infer missing microstructural information from a known composition and processing route. Such process-aware models are critical for predicting properties governed by cross-scale mechanisms that evolve over time, such as creep life, which is driven by the coupled evolution of dislocation motion, grain-boundary sliding and precipitate morphology, all shaped by processing history. Looking further ahead, these process-aware capabilities could be unified within materials-native multimodal foundation models pretrained on heterogeneous data and adapted to diverse downstream tasks, rather than being trained separately for each material class or task.

Building on process-aware representations, generative AI with a feasibility-first approach can propose candidates that are both high-performing and experimentally realisable. The premise is that the bottleneck in materials discovery is increasingly validation rather than candidate generation. In practice, this requires that physical, chemical, processing, and synthesis constraints be embedded in the representation, objective function, or sampling procedure, so that generated candidates remain within experimentally accessible regions. To evaluate feasibility, a generator should return a linked design tuple, comprising a candidate composition or structure, a feasible processing route, an expected microstructural state, target properties under defined testing conditions, and a calibrated uncertainty estimate. Generative models for synthesis planning ~\cite{pan2026diffsyn,prein2025language} are early steps in this direction.
Generated candidates should then undergo staged experimental validation, progressing from computational screening and uncertainty-based ranking to small-batch synthesis and rapid characterisation, with long-duration tests reserved for the most promising candidates. Embedded in a closed-loop workflow, each round of synthesis and characterisation can feed back to refine the generator. This is a form of active learning: rather than testing candidates at random, the workflow prioritises experiments expected to be most informative, such as where the model is least certain, so that each test both filters candidates and improves the generator the most ~\cite{lookman2019active}. Validation should also extend beyond the as-processed state to service-relevant conditions, since deployment novelty depends on performance in service rather than initial properties. Multimodal AI can play an important role in service-life determination, linking processing history, exposure conditions and retained properties to estimate remaining useful life and rank candidates before full-duration testing~\cite{wang2023machine}.

In addition to validating candidates, AI models also need deployment-aware benchmarking that explicitly measures deployment novelty. This benchmarking should rest on three elements: out-of-distribution splits, deployment-relevant metrics, and calibrated uncertainty. Splits such as element holdout, compositional-complexity holdout, processing-route holdout, and time-aware tests aim to push models beyond the training data, but only if the held-out data are genuinely distant along relevant dimensions. Benchmarks should report this distance, so that extrapolation is not mistaken for interpolation. Evaluation metrics should reward deployment performance, such as constraint satisfaction, synthesis success rate, experimental confirmation rate and cost per confirmed candidate, rather than performance scores alone. Uncertainty estimates should be calibrated and tested across modalities, fidelity levels and out-of-distribution settings, using methods such as conformal prediction~\cite{angelopoulos2023conformal} or deep ensembles~\cite{lakshminarayanan2017simple}, so that they remain trustworthy precisely where models are most likely to fail. A deeper and still unresolved question underlies all of these protocols: when data are sparse, it is hard to tell whether a candidate is genuinely new or simply absent from a limited dataset, and no splitting strategy proposed to date fully resolves this.

These four opportunities, multimodal data construction, process-aware modelling, feasibility-first generation, and deployment-aware benchmarking, serve a common goal: grounding higher levels of novelty in stronger evidence. Achieving this requires not only technical progress but also collective action across the materials and AI communities, as discussed next.

\subsection{Key recommendations}

The opportunities above depend on data and the practices that govern how data are reported and shared, which calls for coordinated action across the community. A convening body with standing in the field should draft and maintain a versioned minimum metadata standard that specifies, at minimum, the composition, processing route, a microstructural descriptor, and the measurement conditions for each reported property. Journal editorial boards and funding bodies are well placed to enforce this standard. The former can require that metadata and negative results be deposited alongside reported compositions, much as structural biology made the deposition of atomic coordinates a condition of publication~\cite{berman2003announcing}, albeit with the caveat that materials data are far more heterogeneous and resist a single fixed schema. The latter can attach the same requirements to data management plans, so that the cost of compliant reporting is carried by the grant. To avoid compliance becoming a burden for experimentalists, the community should also provide open, standardised infrastructure, including experimental workflow, metadata templates, data curation tools, and accessible deposition platforms such as GitHub, Hugging Face, or the UK Physical Sciences Data Infrastructure. Agentic AI can further reduce this burden by automating data collection and deposition, turning compliant reporting from a manual overhead into a background process.

Standards and infrastructure only have effect if those who produce and evaluate data act on them. Principal investigators running experimental and self-driving laboratories should report at the sample level, with each measurement linked to its processing and microstructural context, treat failed and null trials as first-class outputs, and, where possible, capture the exposure and post-exposure data needed for service-life prediction. Reviewers should treat the completeness of data and metadata as part of their assessment to ensure reporting standards are consistently enforced in practice. Benchmark maintainers should publish the reference sets and the comparability criteria used to ensure novelty claims to be compared across studies. Taken together, these recommendations provide the opportunities outlined in this section with the evidence they rely on.
}

\section{Conclusion}
\myrev{This Perspective explored whether and how generative and multimodal AI can support the design of novel materials. We pointed out that novelty cannot be reduced to statistical distance in composition or latent space, and we introduced a three-level taxonomy of structural, physical and deployment novelty for evaluating AI-proposed candidates, together with a hierarchy that spans intrinsic properties to extrinsic performance. Seen through this framework, progress has been strong at the structural level, where models now propose diverse and stable candidates at scale, but weak at the physical and deployment novelty levels. The reason is a shortage of evidence: the processing, microstructural and experimental data needed to establish higher-level novelty are scarce, poorly integrated, and evaluated largely against computational proxies.
Closing this gap needs a collective undertaking. The direction we have outlined points towards systems that return not a bare composition or structure, but a complete and experimentally grounded design, one whose feasibility can be tested rather than assumed. Achieving it will require that the data, models, generation, and validation steps be developed together rather than in isolation, and that the evidence on which they depend becomes a shared responsibility of the materials and AI communities. Whether AI can design novel materials, rather than only rediscover known ones, will depend not only on model capacity but also on the evidence we work together to build.}

\newpage

% -----------%
\ack{
X.L., C.A., N.A.M., and H.L. acknowledge funding from EPSRC (grant UKRI396). 
K.A.C, and B.E.J. acknowledge funding from EP/R00661X/1 and EP/S019367/1.
A.J.R acknowledges funding from EPSRC (grant EP/X039285/1).
We thank X. Yang and H. Hassan for collecting related papers.
}

% -----------%
\data{The source data for Figures 1 and 3 are available at https://github.com/multimodalAI/multimodal-ai-landscape/tree/main/data/material-landscape and will be updated annually. An interactive explorer is also available at https://multimodalai.github.io/multimodal-ai-landscape/material.html to facilitate navigation.}

%-----------%
\bibliographystyle{iopart-num}
\bibliography{main}

%-----------%
\newpage
\noindent{\Large\textbf{Supplementary Information}\par}

\section*{S1. Bibliometric analysis of AI for materials research}
\label{sec:supp-s1}

All publication data were retrieved from the Elsevier Scopus database on 24 June 2026 to illustrate recent trends in AI for materials research. Scopus was selected because it provides broad coverage of peer-reviewed literature across materials science, chemistry, physics, engineering, computer science and related disciplines, and supports reproducible keyword-based searches.

All searches were performed using Scopus TITLE-ABS-KEY queries, which match publication titles, abstracts and keywords. Searches were restricted to the Scopus subject areas Materials Science, Computer Science, Chemistry and Engineering, and to the document types Articles, Reviews, Conference Papers and Conference Reviews. The analysis covered the publication years from 2020 to 2025, using the filter \texttt{PUBYEAR > 2019 AND PUBYEAR < 2026}.
The search strategy was based on keyword groups listed in Supplementary Table~\ref{tab:keyword_groups}. Search terms in each group were combined using \texttt{OR}, while different groups were combined using \texttt{AND}. The baseline set for AI in materials research combines the ``AI'' and ``Materials'' keyword groups. All other sets were obtained by adding one or more further keyword groups to this baseline. Annual counts were assigned by publication year. The four data categories (Composition, Microstructure, Processing, and Testing and characterisation) were applied within the ``Multimodal'' group.

Figure~\ref{fig:proportion} shows the proportion of subfields relative to AI for materials publications in the same year: multimodal AI (``AI'' and ``Materials'' and ``Multimodal'' groups), generative AI (``AI'' and ``Materials'' and ``Generative'' groups), and generative and multimodal AI (``AI'' and ``Materials'' and ``Generative'' and ``Multimodal'' groups). Each AI subfield count was divided by the AI for materials count for that year.

Figure~\ref{fig:tasks} shows the task distribution within AI for materials publications, which are property prediction (``AI'' and ``Materials'' and ``Property prediction'' groups) and materials design (``AI'' and ``Materials'' and ``Materials design'' groups). These two task sets are not mutually exclusive, so a publication that mentions both tasks is counted in both, and a publication that mentions neither is counted in neither. Some keywords used in the ``Materials design'' group overlap with those in the ``Generative'' group to help identify a design task. 

Figure~\ref{fig:category-trend} shows AI publications that focus on multimodality and involve at least one of the four data categories. This data was obtained by adding the \texttt{OR} union of all category groups to the ``Multimodal'' group, so that a publication spanning several data categories is counted once, i.e., a deduplicated count. 

Figure~\ref{fig:category-details} shows the relative representation of data categories in multimodal AI publications each year. The base here is the sum of publications across all four categories, rather than the deduplicated count in Fig.~\ref{fig:category-trend}, so a study spanning several categories contributes once to each category. This counting is intentional to show the relationships among categories and the proportions sum to 100\% within each year.

From Fig.~\ref{fig:scopus-mmai} and Fig.~\ref{fig:category}, across 2020 to 2025, AI for materials research grew steadily, and all three AI subfields expanded, with generative AI having the largest share among the three (Fig.~\ref{fig:proportion}). Within this growth, both tasks gained in absolute terms, with property prediction consistently ahead of materials design (Fig.~\ref{fig:tasks}), reflecting that prediction tasks are better established. Multimodal AI also expanded rapidly in the later years (Fig.~\ref{fig:tasks}), and testing and characterisation remained the dominant data category throughout, while microstructure grew over time and composition had the smallest share (Fig.~\ref{fig:category-trend}).

These analyses are inherently approximate and may not capture all relevant studies, particularly those that use different terminology or do not explicitly mention specific methods or data types in the title, abstract or keywords. Therefore, this analysis is intended to illustrate broad trends rather than to provide precise quantitative comparisons.

\renewcommand*{\tablename}{Supplementary Table}
\begin{table}[th]
\centering
\caption{Keyword groups used to construct the Scopus TITLE-ABS-KEY queries. Wildcards follow Scopus syntax: * matches any number of characters and ? matches a single character.}
\label{tab:keyword_groups}
{
\renewcommand{\tabularxcolumn}[1]{>{\raggedright\arraybackslash}m{#1}}
\begin{tabularx}{\textwidth}{>{\raggedright\arraybackslash}m{0.22\textwidth}|>{\raggedright\arraybackslash}X}

\toprule\toprule
\textbf{Keyword groups} & \textbf{Search terms} \\
\midrule
AI & \texttt{AI}; \texttt{machine learning}; \texttt{deep learning}; \texttt{artificial intelligence}; \texttt{graph neural network}; \texttt{neural network}; \texttt{transformer}; \texttt{foundation model}; \texttt{large language model} \\
\midrule
Materials & \texttt{materials science}; \texttt{material*}; \texttt{alloy*}; \texttt{crystal*}; \texttt{high entropy alloy}; \texttt{polymer*}; \texttt{ceramic*}; \texttt{semiconductor*}; \texttt{functional material*} \\
\midrule
Multimodal & \texttt{multimodal}; \texttt{multi-modal}; \texttt{multi modality}; \texttt{multi-modality} \\
\midrule
Generative & \texttt{generative}; \texttt{generative model}; \texttt{generative AI}; \texttt{variational autoencoder}; \texttt{VAE}; \texttt{generative adversarial network}; \texttt{GAN}; \texttt{normalizing flow}; \texttt{energy-based model}; \texttt{energy based model}; \texttt{diffusion model}; \texttt{denoising diffusion} \\
\midrule
Property prediction & \texttt{property prediction}; \texttt{materials property}; \texttt{property model?ing}; \texttt{structure-property}; \texttt{structure property} \\
\midrule
Materials design & \texttt{inverse design}; \texttt{materials design}; \texttt{material design}; \texttt{generative design}; \texttt{de novo design}; \texttt{composition design}; \texttt{structure generation}; \texttt{crystal generation} \\
\midrule
Composition & \texttt{composition}; \texttt{compositional}; \texttt{stoichiometr*}; \texttt{chemical formula}; \texttt{elemental composition}; \texttt{alloy composition}; \texttt{doping}; \texttt{dopant*}; \texttt{atomic fraction}; \texttt{molar fraction} \\
\midrule
Microstructure & \texttt{microstructure}; \texttt{microstructural}; \texttt{grain*}; \texttt{grain boundary}; \texttt{precipitat*}; \texttt{inclusion*}; \texttt{porosity}; \texttt{dislocation*}; \texttt{defect*}; \texttt{texture}; \texttt{crystal structure}; \texttt{atomic structure}; \texttt{crystallographic}; \texttt{crystallography}; \texttt{cif}; \texttt{crystallographic information file}; \texttt{unit cell}; \texttt{lattice parameter*}; \texttt{space group}; \texttt{atomic coordinate*}; \texttt{atomic position*}; \texttt{atomic site*}; \texttt{site occupancy}; \texttt{coordination environment}; \texttt{local atomic environment}; \texttt{atomic arrangement}; \texttt{crystal graph}; \texttt{atomic graph}; \texttt{material graph}; \texttt{structure graph} \\
\midrule
Processing & \texttt{processing parameter*}; \texttt{process parameter*}; \texttt{processing condition*}; \texttt{process condition*}; \texttt{heat treatment}; \texttt{anneal*}; \texttt{quench*}; \texttt{temper*}; \texttt{solution treat*}; \texttt{ageing}; \texttt{aging}; \texttt{sinter*}; \texttt{casting}; \texttt{rolling}; \texttt{forging}; \texttt{extrusion}; \texttt{additive manufacturing}; \texttt{laser powder bed}; \texttt{directed energy deposition}; \texttt{synthesis} AND (\texttt{temperature} OR \texttt{pressure} OR \texttt{time} OR \texttt{cooling rate} OR \texttt{atmosphere}) \\
\midrule
Testing and characterisation & \texttt{characteri?ation}; \texttt{measurement*}; \texttt{testing}; \texttt{microscopy}; \texttt{spectroscopy}; \texttt{diffraction}; \texttt{xrd}; \texttt{sem}; \texttt{tem}; \texttt{ebsd}; \texttt{xps}; \texttt{raman}; \texttt{ftir}; \texttt{eds}; \texttt{edx}; \texttt{tensile}; \texttt{hardness}; \texttt{nanoindentation} \\
\bottomrule\bottomrule
\end{tabularx}
}
\end{table}

\end{document}